\documentclass{nature}

\usepackage{subfigure}
\usepackage{array}
\usepackage{float}

\usepackage{graphicx}
\usepackage{amsmath}
\usepackage{amssymb}
\usepackage{caption}
\usepackage{booktabs}
\usepackage{multirow}
\usepackage{cite}
\usepackage{times}
\usepackage{color}

\title{Analysing Motifs in Multilayer Networks}

\author{Lu Zhong,$^{1}$ Qingpeng Zhang,$^{1\ast}$ Dong Yang,$^{2}$ Guanrong Chen,$^{2}$ Shi Yu $^{3}$ }

\begin{document}

\maketitle

\begin{affiliations}
 \item School of Data Science, City University of Hong Kong, Hong Kong SAR, China
 \item Department of Electronic Engineering, City University of Hong Kong, Hong Kong SAR, China
 \item Department of Neurology, University of Southern California, Los Angeles, CA 90089, US
\end{affiliations}

\begin{abstract}
Network motifs can capture basic interaction patterns and inform the functional properties of networks. However, real-world complex systems often have multiple types of relationships, which cannot be represented by a monolayer network. The multilayer nature of complex systems demands research on extending the notion of motifs to multilayer networks, thereby exploring the interaction patterns with a higher resolution. In this paper, we propose a formal definition of multilayer motifs, and analyse the occurrence of three-node multilayer motifs in a set of real-world multilayer networks. We find that multilayer motifs in social networks are more homogeneous across layers, indicating that different types of social relationships are reinforcing each other, while those in the transportation network are more complementary across layers. We find that biological networks are often associated with heterogeneous functions. This research sheds light on how multilayer network framework enables the capture of the hidden multi-aspect relationships among the nodes.\\
\end{abstract}

\leftline{$\star$ Manuscript Submitted\\}

\clearpage
The increasing complexity of various natural and social systems has changed the communication patterns of multiple agents\cite{cardillo2013emergence,de2016degree,nicosia2015measuring}. Nowadays, people commonly have accounts on different social networking sites and share information within the same site as well as across sites\cite{de2016physics,boccaletti2014structure}. The interactions through the same channel and cross channels can also be observed in, for instance, biological systems\cite{svozil1997introduction,kuurkova1992kolmogorov,battiston2017,Muldoon2016} and transportation systems\cite{gallotti2015multilayer}. In biological systems, protein-protein interactomes have genetic, physical, colocalization and other types of interactions\cite{de2015structural}. Similarly, in transportation systems, multiple modes of transportation (like trains, cars, flights, subways, etc.) enable customers to travel via more flexible and heterogeneous paths\cite{gallotti2015multilayer}. The substantial multiple types rule out the interactions of a traditional monolayer network framework from modelling simultaneous presence and relevance of multiple-mode communications\cite{salehi2015spreading,de2015structural}. Consequently, a multilayer network framework was developed to represent such real-world complex systems, which explicitly incorporates multiple types of agents and multiple modes of interactions\cite{de2016physics,kivela2014multilayer}.

Since the methods and theories on monolayer networks are not applicable to multilayer networks\cite{de2015structural}, their generalisation based on fundamental graph theory became important and necessary, which has already been addressed\cite{wang2006theory,newman2018networks}. Multilayer networks, with multiple types of edges among the same set of nodes, have been used to represent social networks\cite{szell2010multirelational}, gene co-expression networks\cite{li2011integrative}, protein interaction networks\cite{kivela2014multilayer} and transportation networks\cite{battiston2017new}. Studies on multilayer networks found that the structure affects the dynamic process\cite{battiston2017new}, percolation\cite{radicchi2015percolation,callaway2000network} and network evolution. The studies on multilayer networks contributed to the progress of research on interdependent networks and networks of networks\cite{callaway2000network}. Nevertheless, how to characterise natural and engineering network structures has not been well investigated regarding multilayer networks.

Motifs, on the other hand, as the statistically over-represented small connected subgraphs can successfully capture the characteristics of interaction patterns and serve as basic building blocks in monolayer networks\cite{battiston2017,milo2002network}. As a matter of fact, motifs have been used to characterise various families of networks and adopted as a universal design principle for networks\cite{milo2004superfamilies,itzkovitz2005subgraphs}. In a biological network, for example, motifs are basic circuit elements with information-processing functions\cite{alon2007network}. But, whether the monolayer motifs could capture the extremely complicated interaction patterns in multilayer networks is unclear. Several challenging research questions remain: how to define multilayer motifs precisely; what unique functions are processed by multilayer motifs; how are structural features of multilayer motifs related to the functions of the underlying networks\cite{milo2004superfamilies}; what are the topological differences of various networks in social, biological and engineering fields; and so on.

In this paper, we define and extract multilayer subgraphs from some real-world multilayer networks in different fields, i.e., social, biological and transportation networks. We perform statistical analysis on motifs by comparing the distributions of motifs with four identified topological subgraph properties (subgraph density, subgraph triad-density, subgraph correlation, and subgraph cosine-similarity). Analytical results demonstrate that the motifs in social networks and those in a neuronal network have relatively homogeneously distributed subgraph densities and subgraph triad-densities. On the contrary, genetic networks and a transportation network have relatively heterogeneously distributed subgraph densities and subgraph triad-densities. In addition, the motifs are highly correlated and similar across different layers in social networks. Associated with heterogeneous functions, motifs differ each other in subgraph correlation and subgraph cosine-similarity in genetic networks. Finally, motifs in the transportation network were found weakly correlated and dissimilar across layers.

\section*{Results}
\subsection{Multilayer Networks Dataset.} 
\begin{figure*}[t!]
\centering
\includegraphics[scale=0.35]{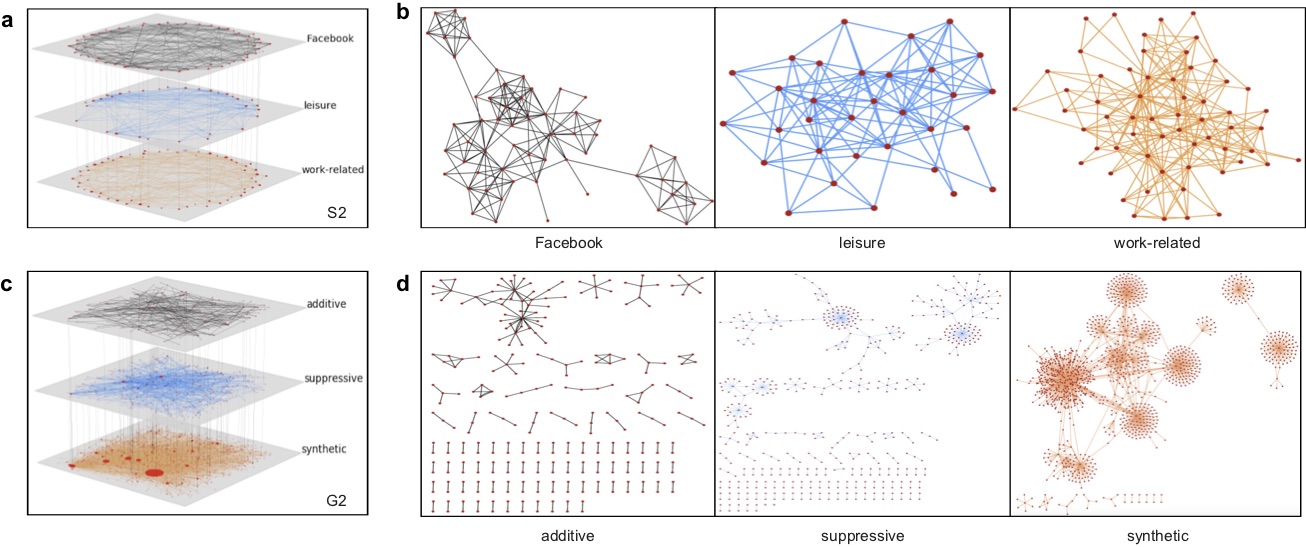}
\caption{The visualisations of a social network ($S2$) and a genetic network ($G_{2}$). (a) the visualisation of the whole multilayer network of $S2$. (b) the visualisations of respective layers in $S2$: the friendships on Facebook (Facebook), interactions during leisure time (leisure), work-related interactions (work-related). (c) the visualisation of the whole multilayer network of $G2$. (d) the visualisations of respective layers in $G2$: additive genetic interactions, suppressive genetic interactions and synthetic genetic interactions.}
\label{visual}
\end{figure*}

In this study, the topological properties of the multilayer motifs in some real-world networks are investigated in three categories (three social networks, four biological networks, one transportation network). Although these networks have different numbers of layers, for consistency only 3-layer networks are selected. Examples of the 3-layer networks are shown in Figure~\ref{visual}. Table 1 summarises the basic statistics and features of these multilayer networks. More details about the original datasets can be found in the Appendix.

\begin{table}
\centering
\caption{Basic statistics of multilayer networks from different datasets. $N$ represents the number of nodes, $M$ represents the number of edges and $L$ represents the number of layers in a network. The elements in $\left \{ \left | e^{1} \right |, \left | e^{2} \right |, ...,\left | e^{\alpha } \right | \right \}$ are the numbers of edges in respective layer $\alpha$. ${M}'$ represents the number of edges with bi-directions.}
\scalebox{0.6}{
\begin{tabular}{|c|c|c|c|c|c|c|c|c|c|}
\hline
            Network & Label    &$L$	&$N$	&$M$	&${M}'$	& $\left \{ \left | e^{1} \right |, \left | e^{2} \right |, ...,\left | e^{\alpha } \right | \right \}$ & Meaning &Type \\
\hline
\multirow{3}*{Social}  &   $S_{1}$  &3 &29 &740 &222  &{361,181,198} &  students  &directed\\  
                   
\cline{2-9}
                     &   $S_{2}$  &3 &61 &553 &0  &{193,124,236}  &  employees &undirected\\  
\cline{2-9}
                     &   $S_{3}$  &3 &71 &2571 &729  &{892,575,1104} &  work associates  &directed\\  
\hline
\multirow{3}*{Genetic}  &   $G_{1}$  &3 &2958 &11864 &956  &{6855,27,4982} &  drosophila melanogaster  &directed\\  
                   
\cline{2-9}
                      &   $G_{2}$  &3 &1772 &2062 &20  &{436,1420,206} &  homo sapiens &directed\\  
\cline{2-9}
                     &   $G_{3}$  &3 &5479 &68495 &6339  &{18629,42361,7505} & saccharomyces cerevisiae &directed\\  
\hline
 Neuronal&   $N_{1}$  &3 &279 &5863 &2758  &{1031,1639,3193}  & caenorhabditis elegans &directed\\  
\hline
 Transportation &   $T_ {3}$  &3 &369 &441 &0  &{312,83,46} & transport in London  &unjpgdirected\\  
\hline
\end{tabular}}
\end{table}

\subsection{Topological Properties of Multilayer Subgraphs.} The abundance of certain types of subgraphs in a given network is related to specific topological properties, such as robustness, stability, assortativity and symmetry. The combinations of subgraphs with different structural properties give rise to the complexity of a network. To unveil combination patterns in multilayer networks, four subgraph properties are employed, which are described in the following.

\emph{Subgraph Density}. It is to measure how densely connected a subgraph is. It equals the number of edges in the multilayer subgraph over the total number of possible edges: 
\begin{equation}
 D_{subgraph}=\frac{\sum_{\alpha }^{L}\sum_{\forall v_{i}, v_{j} \in v} A_{v_{i}, v_{j} }^{\alpha}}{Ln(n-1)}.
\end{equation}

\emph{Subgraph Triad-density}. It measures the transitivity in a subgraph through counting triads. Differing from the transitivity in a monolayer network, nodes form triads through connections in the same layer and also through connections across different layers in a multilayer network. So, the \emph{Subgraph Triad-density} is defined as
\begin{equation}
T_{subgraph}=\frac{\sum_{v_{i} \neq v_{j} \neq v_{k}} {\left | A_{v_{i}, v_{j} }^{\alpha }+A_{v_{j}, v_{i} }^{\alpha } \right |}{\left | A_{v_{i}, v_{k} }^{\alpha }+A_{v_{k}, v_{i} }^{\alpha } \right |}{ \left | A_{v_{j}, v_{k} }^{\alpha }+A_{v_{k}, v_{j} }^{\alpha } \right |}}{\binom{n}{3}{(wL)}^{3}},
\end{equation}
when if the subgraph is undirected, $w=1$; otherwise, if the subgraph is directed, $w=2$.

\emph{Subgraph Correlation}. Real-world networks exhibit multiplex correlations of nodes and their degrees\cite{nicosia2015measuring,lee2012correlated}. Here, a metric is introduced to quantify the interrelationship in a subgraph through counting the presence of edges at different layers:
\begin{equation} 
C_{subgraph}=\frac{\sum_{ \alpha,\beta }P_{\alpha \beta }}{L(L-1)},
\end{equation}
where $P_{\alpha \beta }=\frac{\sum _{v_{i}\neq v_{j}}f(A_{v_{i},v_{j}}^{\alpha },A_{v_{i},v_{j}}^{\beta })}{n(n-1)}$ and $f(A_{v_{i},v_{j}}^{\alpha },A_{v_{i},v_{j}}^{\beta })$ is a piecewise function:
$$f(A_{v_{i},v_{j}}^{\alpha },A_{v_{i},v_{j}}^{\beta })=\begin{Bmatrix}
1, &  A_{v_{i},v_{j}}^{\alpha }=A_{v_{i},v_{j}}^{\beta }=1 ~and ~ A_{v_{j},v_{i}}^{\alpha }=A_{v_{j},v_{i}}^{\beta } ~~~~~~\\ 
0.5, & A_{v_{i},v_{j}}^{\alpha }= A_{v_{i},v_{j}}^{\beta }=1~and ~ A_{v_{j},v_{i}}^{\alpha }\neq A_{v_{j},v_{i}}^{\beta }~~~~~~\\ 
-1, & A_{v_{i},v_{j}}^{\alpha }=A_{v_{j},v_{i}}^{\beta }=1 ~and ~  A_{v_{j},v_{i}}^{\alpha }= A_{v_{i},v_{j}}^{\beta }=0\\
0,& otherwise.
\end{Bmatrix}.$$ 

\emph{Subgraph Cosine-similarity}. Though correlation describes how likely a pair of nodes would been correlated, it could be biased in describing subgraphs for neglecting some topological structures of the whole subgraph. Cosine-similarity is a measure of the angles between two non-zero vectors. If the vectors are independent, then they have a cosine similarity of 0. The \emph{Subgraph Cosine-similarity} is measured by
\begin{equation}
\cos \left ( \theta  \right )=\frac{\sum _{\alpha \neq \beta }\frac{X_{\alpha }X_{\beta }}{\left \|  X_{\alpha }\right \|\left \|  X_{\beta } \right \|}}{L(L-1)},
\end{equation}
where $X_{\alpha}$ is a vector, which is vectorised from $A^{\alpha}$ through concatenating each row into the vector, and similarly $X_{\beta }$ is vectorised from $A^{\beta}$ . 

\begin{figure*}[t!]
\centering
\includegraphics[scale=0.35]{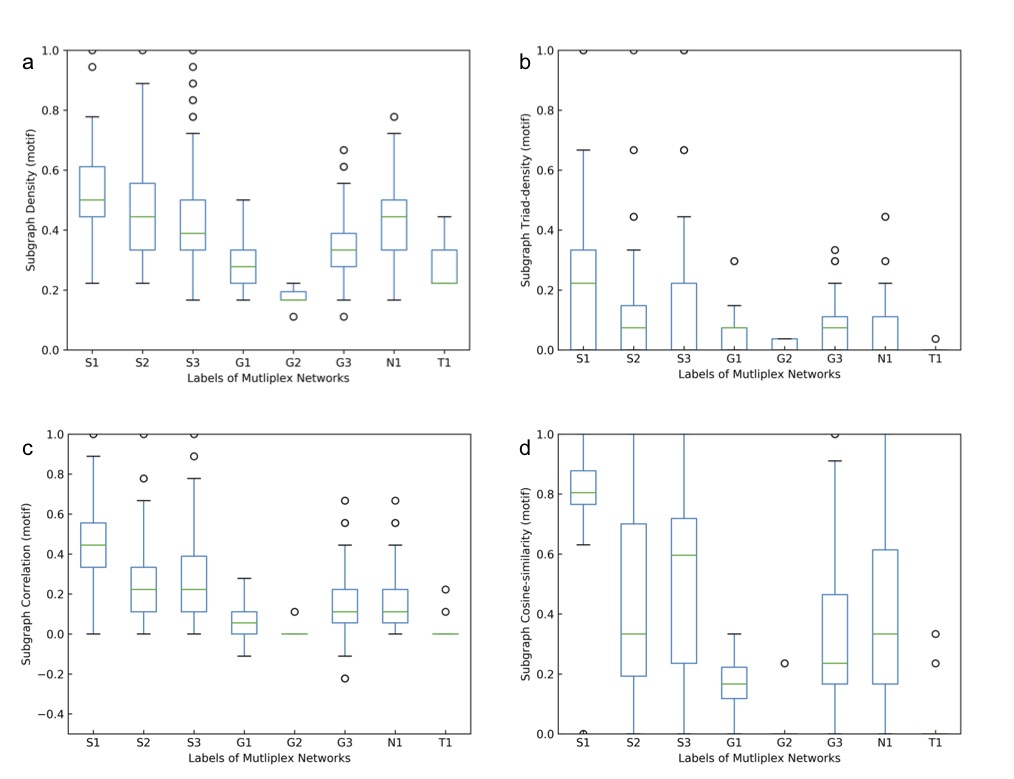}
\caption{Distributions of subgraph density, subgraph triad-density, subgraph correlation and subgraph cosine-similarity of 3-node motifs in multilayer networks. (a) Subgraph Density. (b) Subgraph Triad-density. (c) Subgraph Correlation. (d) Subgraph Cosine-similarity. Reading from bottom up, the box plots include the minimum (at the tip of the lower whisker), first quartile, the median (green line), the third quartile, the maximum (at the tip of the upper whisker), and outliers (green circle). }
\label{AllNet}
\end{figure*}

\begin{figure*}[t!]
\centering
\includegraphics[scale=0.24]{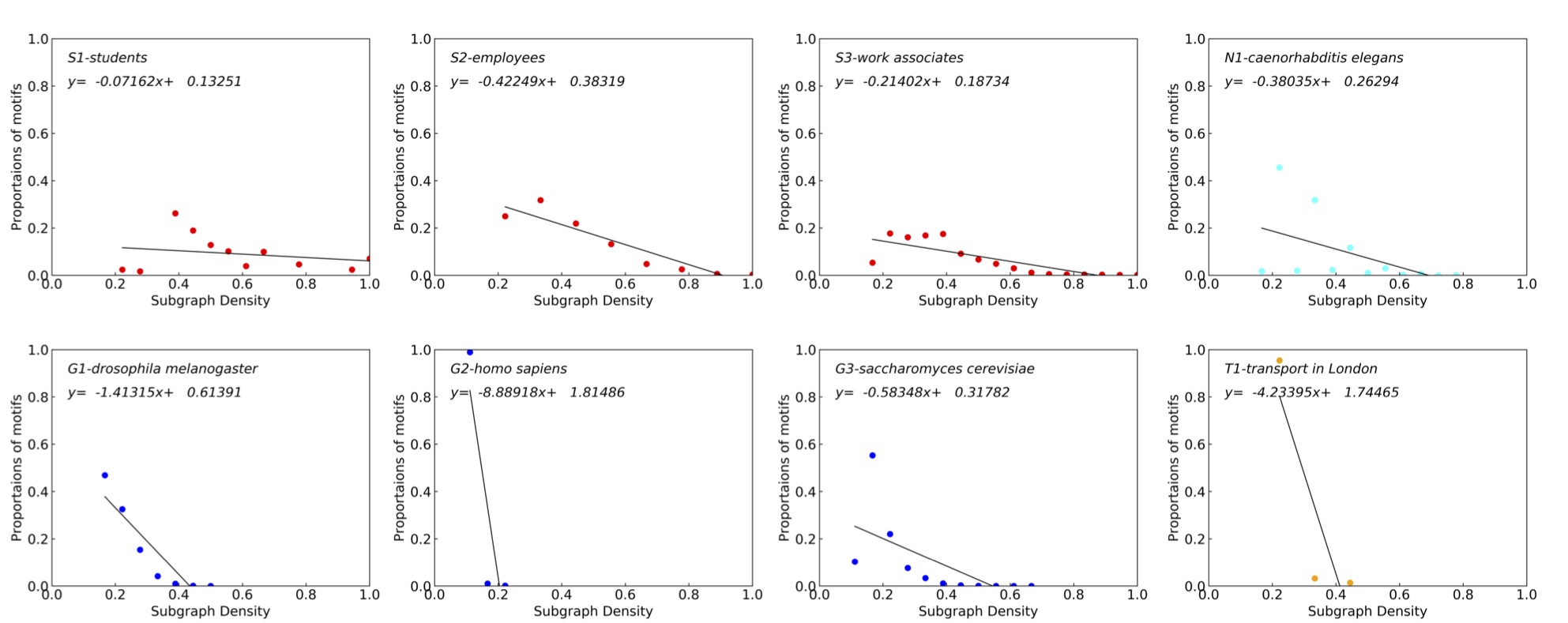}
\caption{Proportions of motifs with subgraph density in multilayer networks.}
\label{density}
\end{figure*} 

\begin{figure*}[t!]
\centering
\includegraphics[scale=0.24]{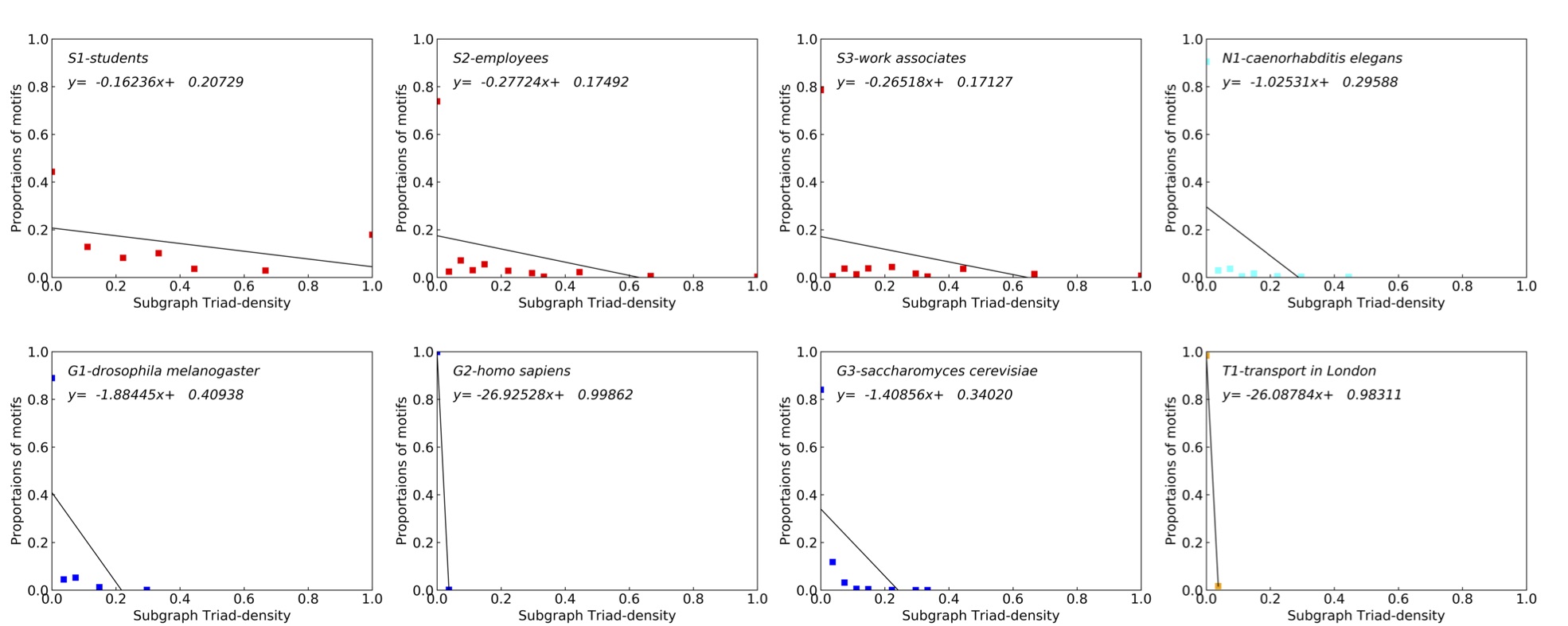}
\caption{Proportions of motifs with subgraph triad-density in multilayer networks.}
\label{triad}
\end{figure*}

\begin{figure*}[t!]
\centering
\includegraphics[scale=0.24]{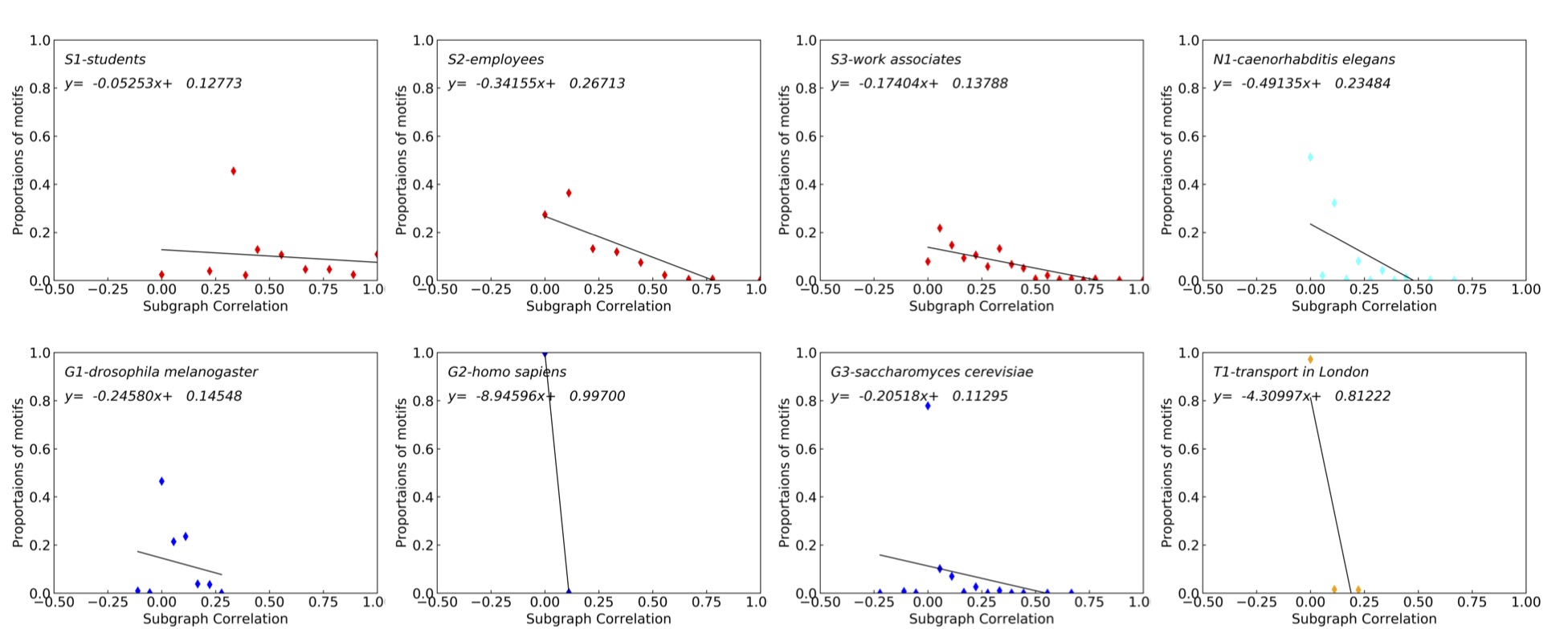}
\caption{Proportions of motifs with subgraph correlation in multilayer networks.}
\label{correlation}
\end{figure*} 

\begin{figure*}[t!]
\centering
\includegraphics[scale=0.24]{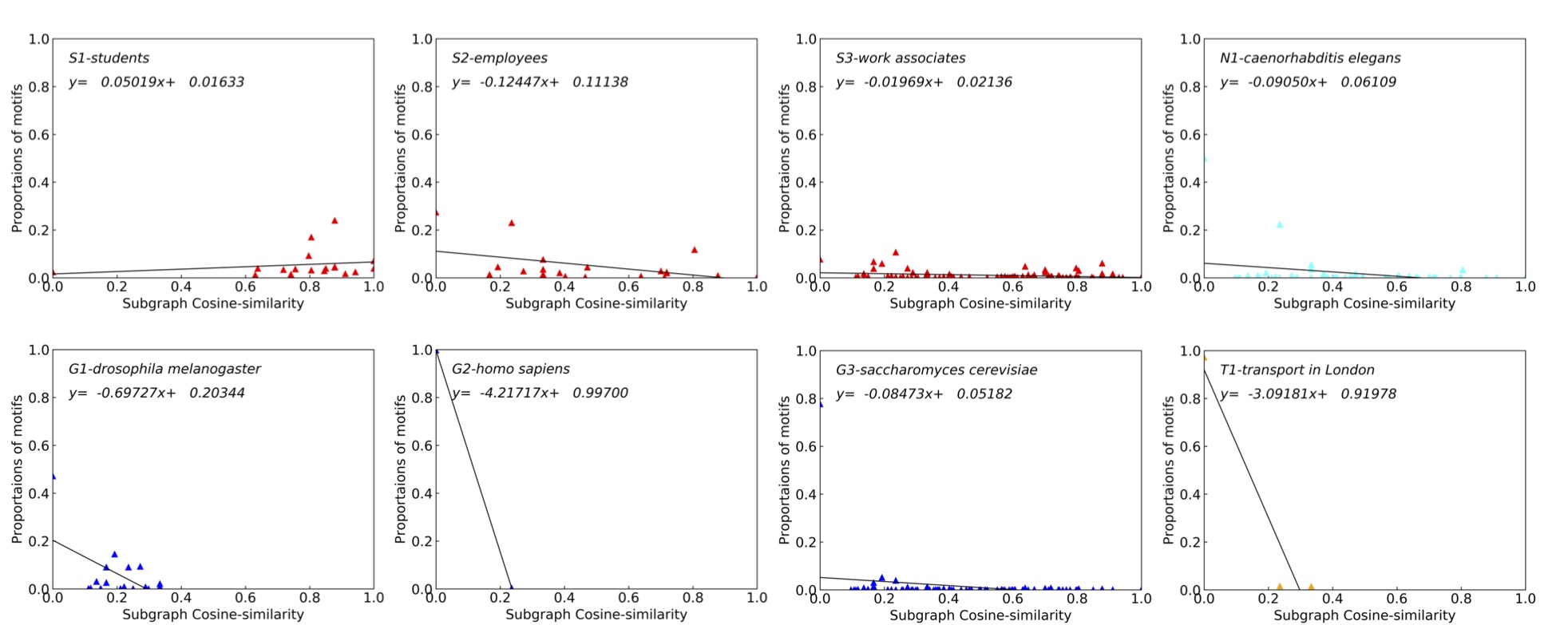}
\caption{Proportions of motifs with subgraph cosine-similarity in multilayer networks.}
\label{similarity}
\end{figure*} 

\begin{figure*}[t!]
\centering
\includegraphics[scale=0.24]{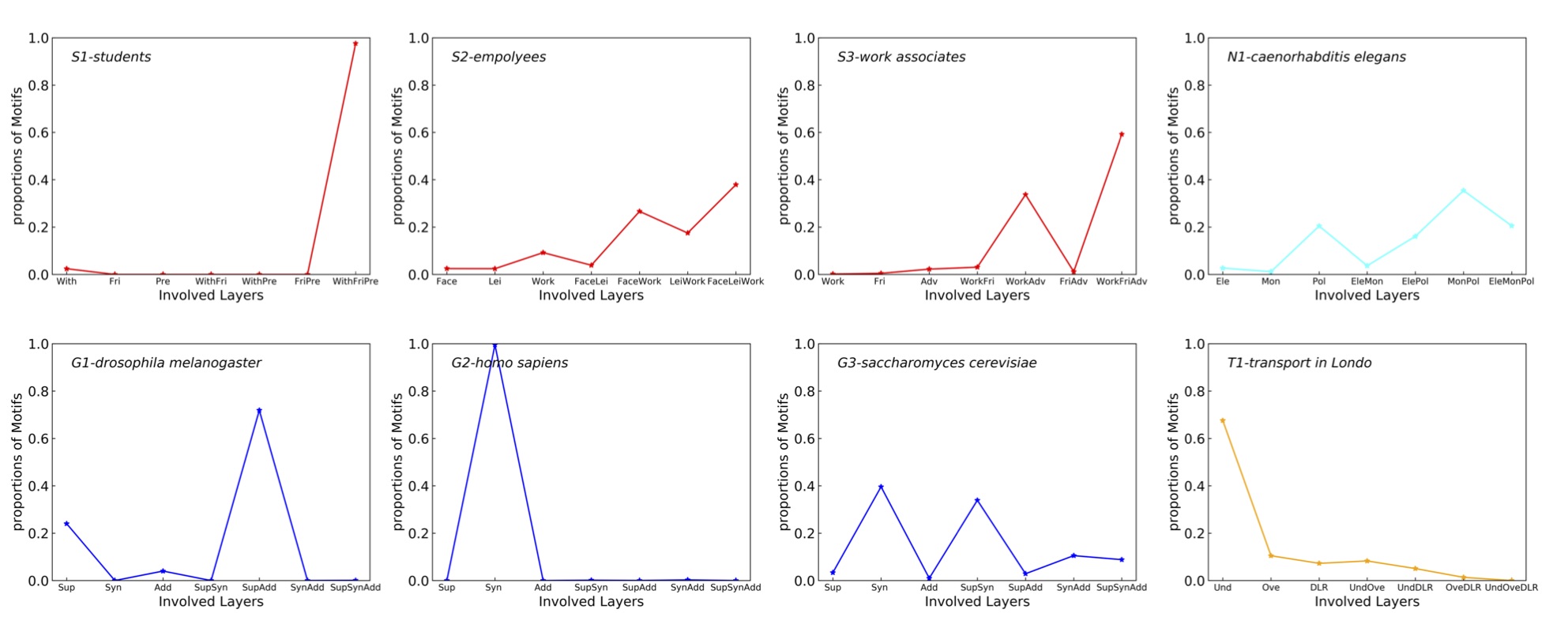}
\caption{Proportions of motifs in involved layers of multilayer networks. Social networks have three layers for the relationships, i.e., getting on with (With), best friendship (Fri) and preference among students (Pre) among students in $S1$; facebook (Face), leisure (Lei), work-related interactions (Work) among employees in $S2$ and co-work (Work), friendship (Fri) and advice relationship (Adv) among work associates in $S3$. The three genetic networks ($G1$, $G2$, $G3$) have same three layers, i.e., suppressive genetic interaction (Sup), synthetic genetic interaction (Syn), and additive genetic interaction (Add). $N1$ represents different types of synaptic junctions among the neurons: electrical (Ele), chemical monadic (Mon), and chemical polyadic (Pol) in caenorhabditis elegans. $T1$ represents three types of transportation in London: underground (Und), overground (Ove) and DLR (DLR).}
\label{layers}
\end{figure*} 

\subsection{Analysis of Multilayer Subgraphs.} Figure~\ref{AllNet} compares the distributions of the above-introduced multilayer motifs with (a) subgraph density, (b) subgraph triad-density, (c) subgraph correlation and (d) subgraph cosine-similarity. Figures 3-6 are the detailed versions of Figure~\ref{AllNet}, which show the proportions of motifs with the four proposed subgraph properties, Figure~\ref{layers} shows the proportions of motifs having edges in different combinations of layers. Through observing these figures, one can draw some general conclusions. The motifs in the examined social networks are denser, more triad-shaped, correlated and topologically similar. On the other hand, motifs in genetic networks ($G1$, $G2$, $G3$) are least dense, less triad-shaped and their topological correlations and similarities across layers are lower. The topological similarities of motifs in $G3$ are higher than other genetic networks, because of the high prevalence of motifs involving both suppressive genetic interactions and synthetic genetic interactions in Figure~\ref{layers}. The motifs in the neuronal network ($N1$) have a similar pattern with that in $S2$. The motifs in the transportation network ($T1$) are less dense, least triad-shaped, lowest topologically correlated and lowest topologically similar. 

More precisely, the motifs in social networks are relatively homogeneously distributed with subgraph density, subgraph triad-density, and subgraph cosine similarity. It is in line with the literature\cite{marsden1988homogeneity, lewis2012social}, where social networks have one common regularity that is the homogeneity. As individuals reinforce their roles by mutually strengthening interactions across layers, the subgraph correlations of motifs are relatively high. The majority of the motifs in $S1$ have high subgraph correlations and they are involved in all three layers. The relationships in $S2$ are least correlated, where the subgraph correlations of the majority of motifs are less than 0.25. As shown in Figure~\ref{layers}, the motifs' distributions of the involved layers in $S2$ and $S3$ share the same pattern, in which the motifs having the work-related relationships are more prevalent. 

\begin{figure*}[t!]
\centering
\includegraphics[scale=0.3]{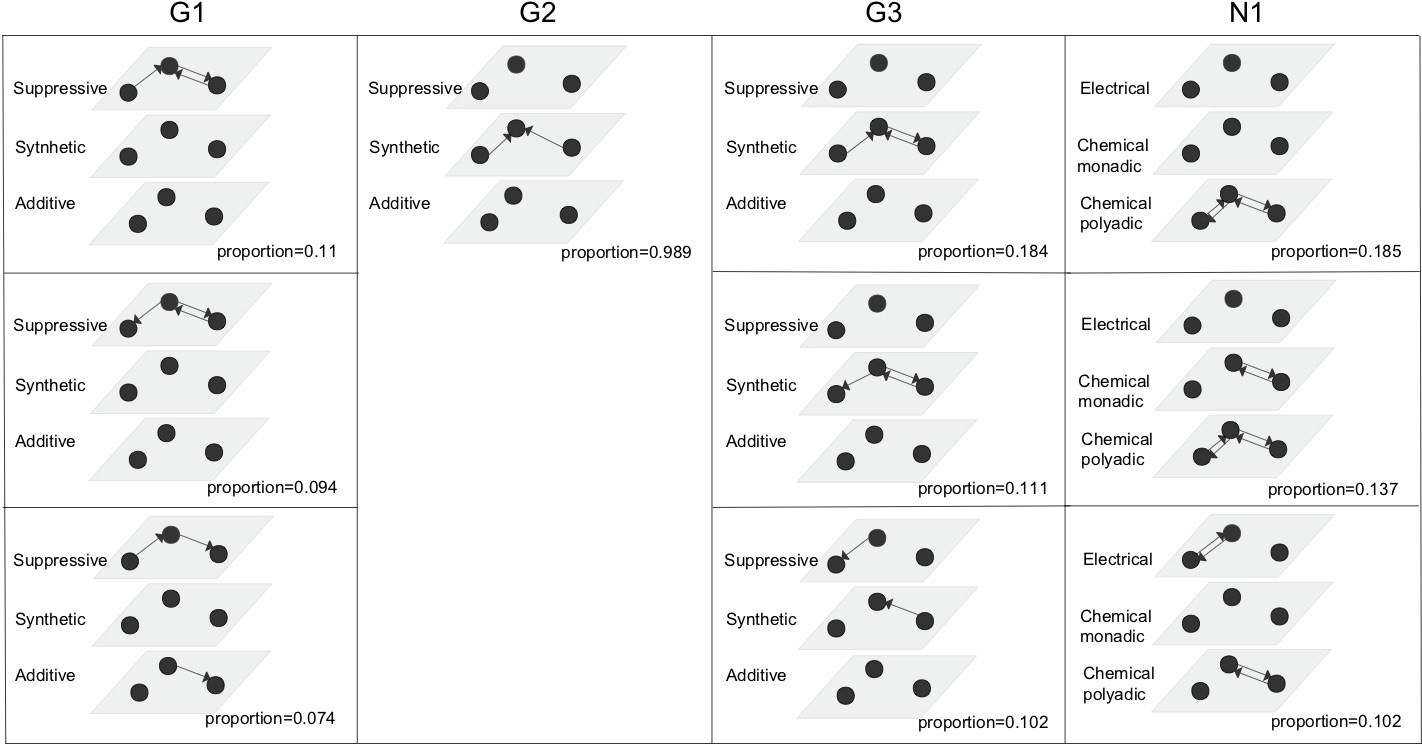}
\caption{Visualisations of motifs occurring frequently in the genetic networks ($G1$, $G2$, $G3$) and the neuronal network ($N1$). The proportion of occurrence of each motif is marked in the bottom right corners of its visualisation. }
\label{MotifExample}
\end{figure*}

Unlike social networks, the motifs in genetic networks are relatively heterogeneously distributed with subgraph density and subgraph triad-density, as shown in Figures~\ref{density},~\ref{triad}. The majority of their motifs are uncorrelated across layers. Corresponding to the differences in functionality, the motifs in the genetic networks differ in both their subgraph cosine-similarities and their involved layers. $G1$ (drosophila melanogaster) has substantial motifs that are low in cosine-similarity Figure~\ref{similarity}. About $70\%$ of motifs in $G1$ involve suppressive genetic interactions and synthetic genetic interactions simultaneously, as shown in Figure~\ref{layers}. The subgraph cosine-similarities of most motifs in $G2$ (homo sapiens) are zero. $G2$ has one dominant motif, which is frequently repeated (98.89\% of all motifs and 42.34\% of all subgraphs). The dominant motif only has synthetic genetic interactions, shaped as $v_{j}\rightarrow v_{i}\leftarrow v_{k}$ in Figure~\ref{MotifExample}. Further investigation on the synthetic genetic interaction in G2 reveals that several nodes are over-represented in the interactions, forming interaction hubs (Figure 1). These hubs are usually associated with specific modules of genetic functions. Interestingly, the interaction partners of those hubs usually do not interact with each other, and hubs in different layers do not overlap much, indicating the existences of modules for diverse functions\cite{van2017gene}. However, it might be due to the limitation of the available genetic interaction for homo sapiens. The motifs in $G3$ (saccharomyces cerevisiae) have high cosine-similarity. In line with the existing research on monolayer networks\cite{milo2004superfamilies}, all the genetic networks here have predominant interaction patterns that transmit information. Figure~\ref{MotifExample} visualises these predominant interaction patterns for each genetic network.

Motifs in the neuronal network are homogeneously distributed with subgraph triad-density, subgraph correlation and cosine-similarity. In particular, both of the subgraph density distribution and subgraph correlation distribution of the neuronal network exhibit high-low turnovers. Three most prevalent motifs are visualised in Figure~\ref{MotifExample}. In general, they share similar patterns: the existence of $v_{j}\rightleftharpoons  v_{i}\rightleftharpoons v_{k}$ in the chemical polyadic layer. Most motifs in $N1$ have more than two types of relationships. Neurons are polarised and have distinct morphological regions with specific physiological functions. Our results suggest that the neurons coexist and cooperate with each other through all relationships\cite{varshney2011structural} to perform their functions.

Like genetic networks, the motifs are heterogeneously distributed in the transportation network. The three modes of transportation in $T1$ are underground, overground (same as the railway) and DLR (Docklands Light Railway). The modes in the transportation network are more complementary to each other, so they typically do not have overlaps except at some transit stations. Most motifs only have edges in the underground layer and the motifs involving multiple modes are rare.

\section*{Discussion}
Multiple relationships commonly exist in real-world complex networks. The existence of multiple relationships results in more complex interaction patterns so that the analysis of multilayer interactions becomes more difficult. The multilayer interaction patterns, which could not be fully captured by the monolayer network framework, demand in-depth analysis by generalising both theories and methodologies to the multilayer network framework. The multilayer network framework offers a much finer resolution to study the complex interaction patterns. This information could be missed from monolayer networks. We may refer to the genetic networks, that the hubs in different layers actually merely overlap, which are interesting and useful insights for the understanding of genetic functions. In this paper, we extend the notion of motifs to multilayer motifs in a comprehensive set of samples ranging from social networks, biological networks (genetic networks and a neuronal network) to a transportation network. Social networks and the neuronal network present densely interconnected triad patterns that correspondingly characterise human relationships and neuronal interactions. Genetic networks and the transportation network present dominant multilayer interaction patterns, which correspondingly involve certain genetic interactions and one transportation mode, respectively. The results shed some light on multilayer interaction mechanisms of social networks, multilayer functional properties of biological networks, and multi-aspect design principles for transportation networks and other engineering networks.

This study has identified intriguing features of interaction patterns in social, biological and transportation networks, leading to new interesting research questions. First, what are the unique information-processing functions processed by the multilayer motifs in biological networks? We observed a number of predominant multilayer interaction patterns in biological networks. Further biological tests are needed to verify the functions associated with the patterns. Second, it is mathematically challenging in enumerating subgraphs and extracting their isomorphic variants. In this paper, we derive formulates to compute the number of all possible 3-node multilayer subgraphs and their node-isomorphic subgraphs. Advanced algorithms and mathematical derivations are needed for $n$-node subgraphs ($n>4$). Third, how to properly apply the empirical and theoretical results of multilayer motifs in the real world? Monolayer motifs have been successfully applied to community detection\cite{porter2009communities}, network design\cite{magnanti1984network}, evolution of biological networks\cite{kashtan2005spontaneous}, percolation of systems\cite{boccaletti2014structure}, and dynamics analysis\cite{waters2012information}. Multilayer motifs have a capability of encoding more information, thus can potentially enhance many applications. This study is limited to eight multilayer network datasets. Some datasets like $G2$ are rather small, simply because the available experimental interaction data were limited. In our ongoing research, we are actively identifying new multilayer motifs with high precision, hoping to make our investigation more complete in the near future.

\section*{Methods}
\subsection{Multilayer subgraph.}In a monolayer network, motifs are small connected subgraphs that are over-represented than the randomised version of the corresponding network\cite{ganter2014predicting,kashtan2004topological}. Subgraphs $g=(v,e)$ in a monolayer network $G=(V,E)$ are classified to various types according to the number of nodes and the placement of edges. In a monolayer network $G$, the smallest subgraphs have $n=\left | v \right |=3$ nodes. These 3-node subgraphs are divided into two groups, namely the chain and triangle subgraphs\cite{ganter2014predicting,estrada2005subgraph}. These two groups of subgraphs have been proved to have significant influence on the topologies and functions of networks, and been applied to many problems like community detection in social networks\cite{evans2010clique} and functional analysis in biological networks\cite{mangan2003structure}.

Extending the notion from single layer to multiple layers, multilayer motifs are small connected multilayer subgraphs that are over-represented in a multilayer network. The multilayer network is defined as\cite{kivela2014multilayer,de2013mathematical} $G^{\alpha }=(V^{\alpha },E^{\alpha })$ with $V^{\alpha }=V$ and $N=\left | V \right |$, where ${\alpha }$ represents a specific layer in ${G^{\alpha }}$. A multilayer subgraph is denoted as $g^{\alpha } =(v^{\alpha },e^{\alpha })$, where $v^{\alpha }=v\in V$ and $n=\left | v \right |$. Furthermore, $g^{\alpha }$ can be classified to various types according to the number of layers, number of nodes and the placement of edges in each layer. Each subgraph could be represented by an adjacency matrix $A=\begin{bmatrix}\begin{smallmatrix}
A^{1} &0  &\cdots  &0 \\ 
0 &A^{2}   & \cdots &0 \\ 
\vdots &\vdots  & \ddots &\vdots  \\ 
0 & 0 & \cdots & A^{L}
\end{smallmatrix}\end{bmatrix}$, where the non-diagonal entries are null matrices and the diagonal entries are the adjacency matrices in each layer of the subgraph.

\subsection{Types of Multilayer Subgraphs.}The smallest multilayer subgraphs ${g^{\alpha }}$ in interest have 2 nodes and 2 layers. The number of possible 2-node (${v_{i},v_{j}}$) multilayer subgraphs is equal to all possible combinations of structures in each layer minus one (the empty case), denoted as
\begin{equation}
N_{2-subgraph}=\omega ^{L}-1,
\end{equation}
where $\omega$ denotes the number of possible connection statuses. In a directed network, $\omega =4$, because there are four possible statuses (null, $\leftarrow$,$\rightarrow$ and $\leftrightarrow$). In an undirected network, $\omega =2$, because there are two possible statuses (null, -). When considering three-node subgraphs it is needed to further account for the connections of different pairs of nodes, thus
\begin{equation}
N_{3-subgraph}=\omega ^{3L}-3(\omega ^{L}-1)-1.
\end{equation}
For a more general case,
\begin{equation}
N_{n-subgraph}=\omega ^{\binom{n}{2}L}- \sum_{i=1}^{n-2}\binom{\binom{n}{2}}{i}(\omega ^{L}-1)^{i}-n \sum_{i=n-1}^{\binom{n-1}{2}}\binom{\binom{n-1}{2}}{i}(\omega ^{L}-1)^{i}-1.
\end{equation}

\subsection{Types of Non-isomorphic Multilayer Subgraphs.}
\begin{figure*}[t!]
\centering
\includegraphics[scale=0.3]{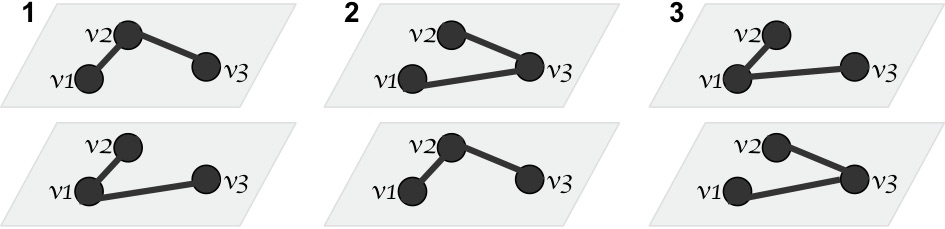}
\caption{An example of three multilayer subgraphs, which are node-isomorphic with each other. Nodes are represented by circles and edge are represented by lines. The edges within the same layer lie in the same shaded parallelogram.}
\label{example}
\end{figure*}

The subgraphs could be node-isomorphic with each other. Node isomorphism refers to that a graph could be mapped to another graph by node-labels permutation without changing the whole structure of each layer\cite{kivela2018isomorphisms}. For two subgraphs ${g^{\alpha }}_{1}$ with node ordering $\pi_{1}$ for adjacency matrix $\mathbf{A_{1}}$ and ${g^{\alpha }}_{2}$ with node ordering $\pi_{2}$ for adjacency matrix $\mathbf{A_{2}}$, there exists a bijective function which rearranges $\pi_{1}$ to $\pi_{2}$ such that eventually $\mathbf{A_{1}}=\mathbf{A_{2}}$. For example, as illustrated in Figure~\ref{example}, although the connections between the same pair of nodes are different, the three subgraphs possess the same adjacency matrix if the nodes are relabelled accordingly. So, these three subgraphs are node-isomorphic with each other. Following the convention, all the node-isomorphic subgraphs are treated as the same subgraph. Counting the non-isomorphic graphs is considered as the graph enumeration problem in graph theory for monolayer networks, and this problem also exists in multilayer networks\cite{kivela2018isomorphisms}. Thus, only enumerations of two-node multilayer subgraphs and three-node multilayer subgraphs are discussed. The enumerations have two parts: undirected subgraphs ($\omega=2$), and directed subgraphs ($\omega=4$). The final numbers of non-isomorphic multilayer subgraphs are listed in Table 2.

When the interactions of two nodes are undirected, the subgraphs are naturally symmetric. Thus, the number of possible two-node multilayer subgraphs, after removing node-isomorphisms, remains the same. When the interactions are directed, simply take half of the sum of all ($4^{L}-1$) subgraphs and the ($2^{L}-1$) subgraphs without node-isomorphic subgraphs with itself:
\begin{equation}
{N}'_{2-subgraph}=\begin{Bmatrix}
2^{L}-1, & \omega =2\\ 
 \frac{2^{L}+4^{L}-2}{2}, & \omega =4
\end{Bmatrix}.
\end{equation}

The number of possible three-node multilayer subgraphs, after removing node-isomorphisms, is not so simple. When $\omega =2$, the number of subgraphs, after removing node-isomorphisms, is $\frac{i^{2}+i^{1}}{2}$ if one edge is fixed. The final number is the sum of counts with all possible fixed edges and then take away the unconnected subgraphs ($\omega =2$): 
\begin{equation}
{N}'_{3-subgraph}=\sum_{i=1}^{2^{L}}\frac{i^{2}+i^{1}}{2}-2^{L}.
\end{equation}
 When $\omega =4$, the enumeration become more complicated and there does not seem to have simple closed form formulates.

\begin{table}
\centering
\caption{Numbers of non-isomorphic multilayer subgraphs. $L$ represents the number of layers in a network and $\omega$ denotes the number of possible connection statuses.}
\scalebox{0.8}{
\begin{tabular}{c|cc|cc}
\hline
~&  \multicolumn{2}{c}{${N}'_{2-subgraph}$}&  \multicolumn{2}{c}{${N}'_{3-subgraph}$} \\
\hline
$L$& $\omega=2$& $\omega=4$& $\omega=2$& $\omega=4$\\
\hline
$1$& $1$& $2$ &$2$ &$13$\\
$2$& $3$& $9$ & $21$&$710$\\
$3$& $7$& $35$ &$112$ &$43932$\\
$4$& $15$& $159 $&$800$&$2798200$\\
\hline
\end{tabular}}
\end{table}

\clearpage
\section*{References}
\bibliography{scibib}

\clearpage
\begin{addendum}
 \item This work was supported by the National Natural Science Foundation of China (NSFC) Grant Nos. 71402157 and 71672163, the Guangdong Provincial Natural Science Foundation No. 2014A030313753, and the Theme-Based Research Scheme of the Research Grants Council of Hong Kong Grant No. T32-102/14N.
 \item[Author contributions] LZ and QZ designed the research; LZ and DY performed the research and analysed the data; LZ, QZ, GC and SY verified, analysed and wrote the paper.
 \item[Competing Interests] The authors declare that they have no competing financial interests.
 \item[Correspondence] Correspondence and requests for materials should be addressed to Qingpeng Zhang.~(email: qingpeng.zhang@cityu.edu.hk).

\end{addendum}

\section*{Appendix }
\subsection{Datasets.}
Here, detailed descriptions of the datasets used in the paper are provided. 

The social networks ($S1$, $S2$, $S3$) capture human’s online or offline interactions, or both. $S1$ is generated by three questions concerning the relationships among 29 seventh grade students in Victoria, Australia\cite{vickers1981representing}. The three questions ask each student to nominate the classmates who are getting on with, the best friends in the class, the classmates who prefer to work with. $S2$ includes both online and offline relationships among the employees of the Computer Science Department at Aarhus\cite{magnani2013combinatorial}. The relationships are friendships on Facebook, interactions during leisure time, work-related interactions (co-authorship and lunch relationships in the raw dataset are integrated as work-related interactions). $S3$ encapsulates Co-work, Friendships and Advice relationships between partners and associates of a Corporate\cite{lazega2001collegial}.

All the genetic networks ($G1$, $G2$, $G3$) are collected from Biological General Repository Interaction Datasets (BIOGRID-3.5.165)\cite{stark2006biogrid} and three layers are extracted from the original datasets. The three layers are suppressive genetic interaction defined by inequality, synthetic genetic interaction defined by inequality, and additive genetic interaction defined by inequality. Genetic interaction could be observed in the relations among the phenotypes of four genotypes: a reference genotype which is the 'wild type' (WT), a perturbed genotype (A), a perturbed genotype (B) and a doubly perturbed genotype (AB). Suppressive interaction ($WT=A=AB<B$) means that A has an effect on WT, but that effect is abolished if a suppressor B is added. Additive interaction ($AB<A<B<WT$) means that two effects are added as a double-mutant effect. Synthetic interaction ($AB<WT<A=B$) means that A and B have no effect but AB has an effect. Suppressive genetic interaction includes dosage rescue, synthetic rescue and phenotypic suppression. Synthetic genetic interaction includes dosage growth defect, dosage lethality, synthetic growth defect, synthetic haploinsufficiency and synthetic lethality. Additive genetic interaction includes phenotypic suppression. Each network refers to different species. $G1$ is about drosophila melanogaster; $G2$ is about homo sapiens; $G3$ is about saccharomyces cerevisiae. 

The neuronal network $N1$ illustrates the caenorhabditis elegans connectome\cite{chen2006wiring}. The three layers are synaptic junctions: electrical, chemical monadic, and chemical polyadic.
 
The three layers in the transportation network $T1$ represents the underground, overground and DLR\cite{de2014navigability}. 

\subsection{Multilayer Motif Detection.}
The detection of multilayer motifs follows the same criteria for detecting the motifs in monolayer networks\cite{milo2002network}. In this paper, we test whether the probability, that a subgraph’s appearance in a randomised network ($N_{rand}$) is equal to or greater than its appearance in the real network ($N_{real}$), is smaller than $P = 0.01$25. Each network is randomised for 1000 times.

The randomised networks are generated by using a randomisation algorithm, which randomises edges in each layer independently through swapping edges. Two edges in the same layer are randomly chosen and then swapped. Swapping edges were repeated until the whole network is well randomised. The randomisation algorithm guarantees that the randomised networks have the same numbers of nodes and edges, and the same degree for each node at each layer, with the original network. 


\end{document}